\documentstyle[preprint,aps,epsf,psfig,epsfig,citesort]{revtex} 
\newif\iftightenlines\tightenlinesfalse
\tightenlines\tightenlinestrue

\newcommand{\oalone}{${\cal{O}} \left( \alpha_s^1 \right)$}
\newcommand{\oalzero}{${\cal{O}} \left( \alpha_s^0 \right)$}

\begin{document}

\title{Tau neutrino deep inelastic charged current interactions}
\author{S.~Kretzer$^1$ and M.~H.~Reno$^2$}
\address{
$^1$Department of Physics and Astronomy, 
Michigan State University, East Lansing, MI 48824 \\
$^2$ 
Department of Physics and Astronomy, University of Iowa, Iowa City,
Iowa 52242}

\maketitle

\begin{abstract}
The $\nu_{\mu} \to \nu_{\tau}$ oscillation hypothesis
will be tested through $\nu_\tau$ production of $\tau$ 
in underground neutrino telescopes
as well as long-baseline experiments.
We provide the full QCD framework for the evaluation 
of tau neutrino deep inelastic charged current (CC)
cross sections, including {\it next-leading-order} (NLO) corrections, 
charm production, tau threshold, and target mass effects in the collinear approximation. 
We investigate
the violation of the Albright-Jarlskog relations for the structure
functions $F_{4,5}$ which occur only in heavy lepton ($\tau$) scattering.
Integrated CC cross sections are evaluated naively over the 
full phase space and with the inclusion of DIS kinematic cuts. 
Uncertainties in our evaluation based on scale dependence, PDF errors
and the interplay between kinematic and dynamical power corrections
are discussed and/or quantified. 
\end{abstract}

\section{Introduction}
\label{sec:intro}

Results from the SuperKamiokande underground experiment measuring the 
atmospheric neutrino flux suggest that muon neutrinos oscillate into 
tau neutrinos with nearly maximal mixing \cite{Fukuda:1998ub}. 
A test of the oscillation
hypothesis is $\nu_\tau$ production of $\tau$
through charged current interactions, a process which will be studied
in underground neutrino telescopes \cite{unt}
as well as long-baseline experiments \cite{lble}
measuring neutrino fluxes from accelerator sources.
For precision measurements of oscillation mixing angles and eventually
CP violation, neutrino cross sections will ideally be know to the level
of a few percent. Eventually, 
measurements of neutrino-nucleon charged current interaction cross sections
are expected to be at the 1\% level at a neutrino factory 
\cite{Albright:2000xi}.

The QCD theory 
of deep inelastic cross sections in the leading-twist approximation
has proceeded to the point of 
evaluating
next-to-next-to leading order (2 loop)
perturbative QCD corrections to
the coefficient functions for the structure functions $F_1,\ F_2$ and $F_3$
\cite{Devoto:wu,Zijlstra:1992kj,Kazakov:fu,Larin:fv}
and approximations for the splitting functions
\cite{Larin:1996wd,Retey:2000nq,vn}.
In the specific case of $\nu_\mu$-isoscalar nucleon ($N$) cross sections,
target mass corrections and nuclear binding effects in a LO twist-2
approach of DIS, combined with the elastic peak 
and a modeling of higher twists in the continuum and resonance region of DIS,
have also been investigated \cite{py,by}.
Tau neutrino charged current interactions with nucleons have received
less theoretical attention \cite{gallagher,py,Dutta:1999jg}. 
Albright and Jarlskog, in Ref.~\cite{aj},
pointed out that there are two additional structure functions, $F_4$ and
$F_5$ that contribute to the tau neutrino cross section. $F_4$ and
$F_5$ are
ignored in muon neutrino interactions because of 
a suppression
factor depending on the square of the charged lepton mass ($m_\ell$)
divided by the nucleon
mass times neutrino energy,
$m_\ell^2/(M_N E_\nu)$. At leading order, in the
limit of massless quarks and target hadrons, $F_4$ and $F_5$ are
\begin{eqnarray} \label{eq:aj1}
F_4 & & =0 \\ \label{eq:aj2}
{2x F_5} & & = F_2 \ ,
\end{eqnarray}
where $x$ is the Bjorken-$x$ variable.
These generalizations of the Callan-Gross relation $F_2=2xF_1$ are
called the Albright-Jarlskog relations.
As with the Callan-Gross relations, the Albright-Jarlskog relations are
violated 
from kinematic mass corrections and 
at NLO\footnote{We will find below that
Eq.~(\ref{eq:aj2}) is {\it not} violated in {\it massless} NLO QCD.}
in QCD. We quantify the violation in Section \ref{sec:sf}.

Below $E_\nu\sim 10$ GeV for muon neutrinos and higher energies for
tau neutrinos, untangling the exclusive and inclusive
contributions to the neutrino-nucleon cross section is difficult.
The cross section in this energy range has quasi-elastic, resonant 
production (such as $\Delta$
production), non-resonant pion production and other inelastic contributions.
There are several 
phenomenological methods for avoiding double counting in this
region. 
One method employs a cutoff in the hadronic invariant mass $W>W_{\min}$
such that the combined exclusive and $W_{\min}$ dependent inclusive
cross section yields a total consistent with data
\cite{Lipari:1994pz}. Using muon neutrino
data, a common choice for $W_{\min}$ is 1.4 GeV. A second method involves
looking at different final state multiplicities \cite{gallagher}. 
By normalizing
the calculated total cross section for a given multiplicity $j$, 
$\sigma_{\rm tot}^j$, to the data, one can determine the factors $f_j$ in
$\sigma_{\rm tot}^j=\sigma_{\rm res}^j+f_j\sigma_{\rm DIS}^j$. 
The quantity $\sigma_{\rm DIS}^j$
is determined by the inelastic cross section and the hadronization
scheme. The Monte Carlo for the Soudan experiment
does the normalization at $E_\nu=20$ GeV 
\cite{gallagher}.

We present here the deep-inelastic contribution to $\nu_\tau N\rightarrow
\tau X$ incorporating next-to-leading order QCD corrections, 
collinear target mass
and charmed quark mass corrections. The NLO structure functions $F_4$ and
$F_5$ including charm quark production have been evaluated by
Gottschalk in Ref.~\cite{gottsch}. Here, we correct a misprint
and evaluate gluonic helicity states in $D$ dimensions rather than 4 
dimensions. 
We show numerical results for the new structure functions and
for the charged current cross sections, both with and
without imposing a nonzero value for $W_{\min}$. To gauge the effects
of perturbative delicacies such as
higher-twist beyond the inclusion of target mass effects 
in the scaling variable,
we also evaluate the effect of imposing a cutoff on
$Q^2$. We find that the collinear cutoff effect of
the tau mass $m_{\tau}=1.78\ {\rm GeV}$ is enough to guarantee
that for $E_{\nu_{\tau}} \gtrsim 5\ {\rm GeV}$,  
$\sigma_{\rm CC}({\nu_{\tau}}N)$ is dominated by $Q^2 > 1\ {\rm GeV}^2$. While
this is widely considered a perturbative scale in QCD, some caveats about
the possible importance of higher twist are discussed.

In Section \ref{sec:formulae}, we show the formulae for the differential cross section
in terms of structure functions, and the expressions for the structure
functions at NLO. Charm quark mass corrections for the charged current
process are displayed, together with the results for the $m_c\rightarrow 0$
limit. In Section \ref{sec:sfandxs}, we exhibit our numerical
results for the structure functions $F_4$ and $F_5$, and for
$\nu_\tau N$ and $\bar{\nu}_\tau N$ interactions for
neutrino energies up to 100 GeV. Cross sections are compared to those
with incident muon neutrinos and uncertainties
are discussed. We conclude in Section \ref{sec:sum}.
In an Appendix we rewrite the Albright-Jarlskog relations in terms
of helicity amplitudes and pinpoint the approximations in their 
derivation.

\section{Formulary: $\nu_\tau$ DIS at NLO}
\label{sec:formulae}

Neglecting neither the target nucleon
mass $M_N$ nor the final state lepton mass $m_{\tau}$,
the charged current $\nu_{\tau}$ (anti-)neutrino 
differential cross section is
represented by a standard set of 5 structure functions \cite{aj}\footnote{
Our normalization of $F_4$ differs from that in \cite{aj} 
by a factor of $x$.}
\begin{eqnarray} \nonumber
\frac{d^2\sigma^{\nu(\bar{\nu})}}{dx\ dy} &=& \frac{G_F^2 M_N
E_{\nu}}{\pi(1+Q^2/M_W^2)^2}\
\left\{
(y^2 x + \frac{m_{\tau}^2 y}{2 E_{\nu} M_N})
F_1^{W^\pm} \right. \\ \nonumber
&+& \left[ (1-\frac{m_{\tau}^2}{4 E_{\nu}^2})
-(1+\frac{M_N x}{2 E_{\nu}}) y\right]
F_2^{W^\pm}
\\ \nonumber
&\pm& 
\left[x y (1-\frac{y}{2})-\frac{m_{\tau}^2 y}{4 E_{\nu} M_N}\right]
F_3^{W^\pm} \\  \label{eq:nusig}
&+& \left.
\frac{m_{\tau}^2(m_{\tau}^2+Q^2)}{4 E_{\nu}^2 M_N^2 x} F_4^{W^\pm}
- \frac{m_{\tau}^2}{E_{\nu} M_N} F_5^{W^\pm}
\right\}\ .
\end{eqnarray} 
where $\{x,y,Q^2\}$ are the standard DIS kinematic variables 
related through
$Q^2 = 2 M_N E_\nu x y$ and where we have neglected factors of 
$m_\tau^2/2M_N E_\nu\cdot Q^2/M_W^2$.
These latter corrections from the 
$\sim q^{\mu} q^{\nu} /M_W^2$ part of the  
massive boson propagator
are negligible both at low and at high neutrino energies 
and will not enter our numerics.
For completeness of this formulary, though, they
can be included multiplicatively by replacing:
\begin{equation}
F_i^{W^\pm} \rightarrow F_i^{W^\pm} \times \left( 1 + \epsilon_i\right)
\end{equation}
with
\begin{eqnarray} \nonumber
\epsilon_1 &=& 
\frac{m_{\tau}^2\ (Q^2+2M_W^2)}{2 M_W^4}
\\ \nonumber
\epsilon_2 &=& 
-\frac{E_{\nu}^2\ m_{\tau}^2\ y\ [ 4 M_W^2+y (Q^2+m_{\tau}^2)]}{M_W^4\ [4(y-1)E_{\nu}^2+m_{\tau}^2+Q^2]}
\\ 
\epsilon_3 &=& 0 \\ \nonumber
\epsilon_4 &=& 
\frac{Q^2\ (Q^2+2 M_W^2)}{M_W^4}
\\ \nonumber
\epsilon_5 &=&
\frac{Q^2}{M_W^2}+\frac{(M_W^2+Q^2)\ (m_{\tau}^2+Q^2)\ y}{2 M_W^4}
\end{eqnarray}

Kinematics determine the integration ranges\footnote{A typographic
error in Refs.~\cite{aj,py} is corrected here in the lower limit for $x$.}
to be \cite{aj,py}
\begin{equation}
\label{eq:limitx}
{m_\tau^2\over 2 M_N (E_\nu-m_\tau )}\leq x\leq 1
\end{equation}
and 
\begin{equation}
\label{eq:limity}
a-b\leq y \leq a+b
\end{equation}
where
\begin{eqnarray}
\label{eq:limab}
a & & = {1-m_\tau^2\Biggl({1\over 2M_N E_\nu x}+{1\over 2 E_\nu^2}\Biggr)
\over 2\Biggl(1+{M_N x\over 2 E_\nu}\Biggr)} \\ \nonumber
b & & = {\sqrt{\Biggl(1-{m_\tau^2\over 2 M_N E_\nu x}\Biggr)^2-{m_\tau^2\over
E_\nu^2}}
\over 2\Biggl(1+{M_N x\over 2 E_\nu}\Biggr)} \ .
\end{eqnarray}

In the perturbative regime we
can calculate the structure functions $F_i^{W^\pm}$
from parton dynamics by use of the factorization theorem
\begin{equation}
\label{eq:fac}
W^{\mu\nu} = \int\ \frac{d\xi}{\xi}\ f(\xi,\mu^2)\ 
{\hat{\omega}}^{\mu\nu}|_{p^+=\xi P_N^+}\ \ \ .
\end{equation}
relating the hadronic ($W^{\mu \nu}$)
and partonic (${\hat{\omega}}^{\mu\nu}$)
forward matrix element 
of the product of weak currents $< J^{\mu} J^{\nu} >$. 
In Eq.~(\ref{eq:fac}), $f(\xi,\mu^2)$ is a parton distribution function 
evaluated at factorization scale $\mu$
and the parton momentum fraction $\xi$ of the light cone momentum of the 
nucleon  
$P^+_N \equiv (P^0_N + P^z_N )/\sqrt{2}$. Intrinsic transverse
momentum of the incoming parton is neglected throughout our discussion,
{\it i.e.}~$p_\perp =0$ in (\ref{eq:fac}).

For the neutrino energies of interest here, we can safely
restrict the consideration to the first two quark generations.
The light flavour contributions to Eq.~(\ref{eq:nusig}) can
be obtained from the $m_c \rightarrow 0$ limit of the
charm production component which we will, therefore, consider first.
The charm production contribution to $d \sigma$ will be represented by
the structure functions $F_i^c$. We introduce 
theoretical structure functions\footnote{To meet with an 
experimentalist's point of view,
we denote functional dependence on $x$ and $Q$ only in the LHS of
Eq.~(\ref{eq:conv}); corresponding in theory to an exact knowledge 
of all mass parameters and an ideal validity of the renormalization group 
for observable cross sections. Residual scale ($\mu$)
dependence will be investigated below.
} 
\begin{eqnarray}
{\cal{F}}_i^c(x,Q^2)\ \  
=\ \ (1-\delta_{i4})\cdot s^{\prime}({\bar \eta},\mu^2) &+&
\frac{\alpha_s(\mu^2)}{2\pi}\left\{\int_{{\bar \eta}}^1 \frac{d\xi^{\prime}}{\xi^{\prime}}
\left[H_i^q\left(\xi^{\prime},\frac{Q^2}{\mu^2},\lambda\right)
\ s^{\prime}(\frac{{\bar \eta}}{\xi^{\prime}},\mu^2)
\right. \right. \nonumber\\
&+& \left. \left. H_i^g\left(\xi^{\prime},\frac{Q^2}{\mu^2},\lambda\right)
\ g^{\prime}(\frac{{\bar \eta}}{\xi^{\prime}},\mu^2)
\right]\right\}
\label{eq:conv}
\end{eqnarray}
for scattering off the CKM-rotated weak eigenstate
\begin{equation}
s^{\prime} = \left|V_{s,c}\right|^2\ s + \left|V_{d,c}\right|^2\ d 
\end{equation}
and its QCD evolution partner
\begin{equation} 
g^{\prime}\equiv  g \left(
\left|V_{s,c}\right|^2 + \left|V_{d,c}\right|^2 \right)   
\end{equation}
{\it i.e.}, $d s^{\prime} / d \ln Q^2
= s^{\prime} \otimes P_{qq} + g^{\prime} \otimes P_{qg}$. 

In Eq.~(\ref{eq:conv}) we set the renormalization scale
equal to the factorization scale and
\begin{equation}
{\bar \eta} = \frac{\eta}{\lambda} \ \ \ \ ,
\ \frac{1}{\eta} = \frac{1}{2 x} + \sqrt{\frac{1}{4 x^2} + \frac{M^2}{Q^2}} 
\label{eq:eta}
\end{equation}
is the target mass corrected slow rescaling variable.The quantity $\eta$
is the Nachtmann variable \cite{nacht}. 
The charm mass
dependence is included in the dimensionless 
$\lambda\equiv Q^2/(Q^2+m_c^2)$. In Eq.~(\ref{eq:conv}) the convolution
variable $\xi$ in Eq.~(\ref{eq:fac}) has been traded for
$\xi^{\prime}={\bar \eta}/\xi$ which relates to the 
partonic CMS energy ${\hat s}=(p+q)^2$ 
through $1/\xi^{\prime}=\lambda (1+{\hat s}/Q^2)$.

The theoretical structure functions in Eq.~(\ref{eq:conv}) 
are obtained from a tensor projection on the partonic
${\hat{\omega}}^{\mu\nu}$ and have been 
conveniently normalized to a simple ${\cal{O}}(\alpha_s^0 )$ term. 
As indicated, the leading order contribution to ${\cal F}_4^c$ vanishes.
In the presence of target mass,
the physical structure functions $F_i^{W^\pm}$ in Eq.~(\ref{eq:nusig}) 
- relating to a tensor projection on the hadronic $W^{\mu \nu}$ -
are a mixture\footnote{See Ref.~\cite{aot} for details on the mixing of
structure functions within the tensor basis.} 
of ${\cal{F}}_i$. Explicitly
in our case of CC charm production we have:
\begin{eqnarray} \label{eq:mix1}
F_1^c &=& {\cal{F}}_1^c \\ \label{eq:mix2}
F_2^c &=& 2\ \frac{x}{\lambda}\ \frac{{\cal{F}}_2^c}{\rho^2} \\ \label{eq:mix3}
F_3^c &=& 2\ \frac{{\cal{F}}_3^c}{\rho}\\ \label{eq:mix4}
F_4^c &=& \frac{1}{\lambda }\ \frac{(1-\rho )^2}{2 \rho^2 }
\ {\cal{F}}_2^c + {\cal{F}}_4^c + \frac{1-\rho}{\rho}\ {\cal{F}}_5^c \\ \label{eq:mix5}
F_5^c &=& \frac{{\cal{F}}_5^c}{\rho} - \frac{(\rho -1)}{\lambda \rho^2}
\ {\cal{F}}_2^c
\end{eqnarray} 
where 
\begin{equation}
\rho^2 \equiv 1 + \left( \frac{2 M_N x}{Q}\right)^2 \ ,
\end{equation}
$F_i^c=F_i^c(x,Q^2)$ and 
${\cal{F}}_i^c={\cal{F}}_i^c(x,Q^2)$. 
These results are in agreement with Table V in Ref.~\cite{aot}. They
rely only on the assignment of parton light cone momentum $p^+$ related
to the nucleon light cone momentum $p^+=\xi P^+_N$, with $p_\perp=0$ and
$\xi=\bar{\eta}$ at leading order.

The NLO corrections $H_{i=1,...,5}^{q,g}$ were first obtained in 
Ref.~\cite{gottsch} and $H_{1,2,3}^{q,g}$ have been
rederived in \cite{gkr}. We follow these references in notation and
present the full set  $(i=1,...,5)$ including   
our independently rederived $H_{4,5}^{q,g}$. 
The fermionic NLO coefficient functions $H_i^q$ 
in Eq.~(\ref{eq:conv}) for charm
production are calculated from the subprocess $W^+ s
\rightarrow g c$ and from the one-loop correction to $W^+ s \rightarrow c$. 
They are given by\footnote{For the ease of notation, from here
on $\xi$ will be a generic variable and no longer the light-cone
momentum of Eq.~(\ref{eq:fac}). To strictly match Eq.~(\ref{eq:conv})
we should have $\xi \rightarrow \xi^{\prime}$ instead.}
\begin{equation}
H_{i={1,2,3,5}}^q\left(\xi,\frac{Q^2}{\mu^2},\lambda\right)\ =\ \left[ P_{qq}^{(0)}(\xi)\
\ln\frac{Q^2}{\lambda\ \mu^2}\ +\  h_i^q(\xi,\lambda)\right]
\label{qincupol}
\end{equation}
where $\displaystyle \quad
P_{qq}^{(0)}(\xi)=\frac{4}{3}\left(\frac{1+\xi^2}{1-\xi}\right)_+ \quad$ 
and
\begin{eqnarray}
h_i^q(\xi,\lambda)\ =\ \frac{4}{3} &\bigg\{& h^q+A_i\ \delta
(1-\xi)+B_{1,i}\ \frac{1}{(1-\xi)_+} \nonumber\\
 &+& \left. B_{2,i}\ \frac{1}{(1-\lambda
\xi)_+}+B_{3,i}\ \left[\frac{1-\xi}{(1-\lambda \xi)^2}\right]_+\right\}
\label{eq:smallhq}
\end{eqnarray}
with
\begin{eqnarray}  \nonumber
h^q\ =\ &-&\left(4+\frac{1}{2\lambda}+\frac{\pi^2}{3}+\frac{1+3\lambda}
{2\lambda}\ K_A\right)\delta(1-\xi) \\
 &-& \frac{(1+\xi^2)\ln \xi}
{1-\xi}+(1+\xi^2)\left[\frac{2\ln (1-\xi)-\ln (1-\lambda \xi)}{1-\xi}\right]_+
\end{eqnarray}
and
\begin{equation}
K_A\ =\ \frac{1}{\lambda}\ (1-\lambda)\ \ln (1-\lambda)\ \ \ .
\label{ka}
\end{equation}
The coefficients in (\ref{eq:smallhq}) for $i=1,2,3,5$ are given in Table \ref{thq}. 
A misprint in Ref.~\cite{gottsch} concerning $A_2$ was already 
corrected in \cite{gkr}. 
Here, we correct a similar misprint concerning  $A_5$.

The gluonic NLO coefficient functions $H_i^g$ 
in Eq.~(\ref{eq:conv}) for charm
production, as calculated from the subprocess $W^+ g
\rightarrow c \bar{s}$, are given by
\begin{equation}
H^g_{i={1,2,5 \atop 3}}\left(\xi,\frac{Q^2}{\mu^2},\lambda\right)\ =\
\left[P_{qg}^{(0)}(\xi)\left(\pm L_{\lambda}+
\ln\frac{Q^2}{\lambda\ \mu^2}
\right)+h_i^g(\xi,\lambda)\right]
\label{gincupol}
\end{equation}
where $\displaystyle \quad
P_{qg}^{(0)}(\xi)\ =\ \frac{1}{2}\ \left[ \xi^2+(1-\xi)^2 \right],
\quad L_{\lambda}\ =\ \ln\frac{1-\lambda \xi}{(1-\lambda)\xi} \quad$ and
\begin{equation}
h_i^g(\xi,\lambda)\ =\ C_0+C_{1,i}\ \xi(1-\xi) + C_{2,i} + (1-\lambda)\ \xi
\ L_{\lambda}\ (C_{3,i}+\lambda\ \xi\ C_{4,i})
\label{hig}
\end{equation}
with
\begin{equation}
C_0\ =\ P_{qg}^{(0)}(\xi)\ \left[2\ln (1-\xi)-\ln (1-\lambda \xi) -\ln
\xi\right]\ \ \ .
\end{equation}
The coefficients $C_{k,i}$ are given in Table \ref{thg}. 
As in Ref.~\cite{gkr},
they differ from those in
Ref.~\cite{gottsch} by counting 
- within dimensional regularization - 
the gluonic helicity states in 
$D$ \cite{bbdm,fupe} rather than in $4$
dimensions.

The structure function $F_4$ is insensitive
to any collinear physics up to ${\cal{O}}(\alpha_s^1 )$
and the corresponding coefficients
$H_{i=4}^{q,g}$ are, therefore,
scheme and scale-independent functions. They are given by
\begin{equation}
H_{i=4}^q (\xi, \lambda) =
\frac{4}{3}\ \frac{\lambda (1-\xi ) \xi \left[1+(1-2 \lambda )\xi \right]}
{(1-\lambda \xi)^2}
\end{equation}
\begin{equation}
H_{i=4}^g (\xi, \lambda) =
2 \lambda \xi \left[
1-\xi - (1-\lambda ) \xi\ L_{\lambda} \right]\ . 
\end{equation}

In the limit $\lambda\rightarrow 1\ (m_c\rightarrow 0)$ 
and after an additional minimal subtraction of the
collinear mass singularities in $H_{i=1,2,3,5}^g$
the $H_i^{q,g}$ reduce to the 
massless $\overline{\rm{{MS}}}$ coefficient 
functions:\footnote{The choice $\mu=Q$ is made to match with the notation in \cite{fupe}.
Retaining a  general $\mu \neq Q$ on the LHS of (\ref{eq:limq}), (\ref{eq:limg}) results in
the massless coefficient functions for arbitrary scale.
Obviously, this amounts to adding back the
$P^{(0)} \ln (Q^2/\mu^2)$
``splitting function times log'' counter-terms.}
\begin{eqnarray} \label{eq:limq}
\left.
\lim_{\lambda\rightarrow 1} H_i^q\left(\xi,\frac{Q^2}{\mu},\lambda\right)
\right|_{\mu^2=Q^2}
&=&  C_{F,i}^{(1)}(\xi)\\ \label{eq:limg}
\left.
\lim_{\lambda\rightarrow 1} \left\{
H_{i={1,2,4,5 \atop 3}}^g\left(\xi,\frac{Q^2}{\mu^2},\lambda\right)
\mp   (1-\delta_{i4}) P_{qg}^{(0)} \ln \frac{\mu^2/Q^2}{1- \lambda } 
\right\}\right|_{\mu^2=Q^2}
&=& C_{G,i}^{(1)}(\xi)
\end{eqnarray} 
where $C_{F,i}^{(1)}$ and $C_{G,i}^{(1)}$ 
for $i=1,2,3$ are the massless $\overline{\rm{{MS}}}$ coefficients in
Appendix III of \cite{fupe}. 
Extending the results in Refs.~\cite{fupe,aem,bbdm} to include $i=4,5$
and within the notation of Ref.~\cite{fupe} we find 
\begin{eqnarray} \label{eq:cf4}
C_{F,4}^{(1)}(\xi)
&=& \frac{4}{3}\ \xi
\\  \label{eq:cf5}
C_{F,5}^{(1)}(\xi)
&=& C_{F,2}^{(1)}(\xi)
\\  \label{eq:cg4}
C_{G,4}^{(1)}(\xi)
&=& 2\ \xi\ (1-\xi )
\\  \label{eq:cg5}
C_{G,5}^{(1)}(\xi)
&=&  C_{G,2}^{(1)}(\xi)\ .
\end{eqnarray}
To calculate the light quark contributions to CC $\nu_{\tau}$ DIS,
the massless coefficient functions on the RHS of 
Eqs.~(\ref{eq:limq})-(\ref{eq:cg5})
are used together with the PDFs which multiply 
the CKM matrix elements $\left|V_{i,j\neq c}\right|^2$  
in an obvious modification ($\lambda \rightarrow 1$, $H\rightarrow C$, ...)
of our Eq.~(\ref{eq:conv}). 
The result is added to the charm production 
component (\ref{eq:conv}) to obtain the entire NLO structure function.

It is interesting to note that
the equalities (\ref{eq:cf5}), (\ref{eq:cg5}) guarantee that
the Albright-Jarlskog relation (\ref{eq:aj2}) is not violated in
massless QCD at NLO. Eqs.~(\ref{eq:mix1})-(\ref{eq:mix5})
and the fact that $H_2^{q,g} \neq H_5^{q,g}$ for charm production manifest, of course,
a violation of Eq.~(\ref{eq:aj2}) in the real world of massive target hadrons and 
of heavy quarks interacting through QCD. This observation must actually be
expected to hold at all orders as we clarify in the Appendix.

\section{Structure Functions and Cross Sections}
\label{sec:sfandxs}

\subsection{Structure Functions}
\label{sec:sf}

The Albright-Jarlskog relations (Eqs.~(\ref{eq:aj1}) and (\ref{eq:aj2}))
are valid at leading order in the massless limit. Here, we show $F_4$ and
$2xF_5-F_2$ to demonstrate violations of these relations. We use the
CTEQ6 parton distributions \cite{Pumplin:2002vw} 
which include estimates of the uncertainties in the distributions. 
Thus, in Section \ref{sec:xsec}
we will be able to quantify the error in the evaluation
of $\sigma_{\rm CC}({\nu_{\tau}}N)$ that is caused by our imperfect
knowledge of the PDFs.
The CTEQ6 fits are provided for $\mu>\mu_0=1.3$ GeV 
but can be extrapolated to lower values of $\mu$.
When the factorization scale squared goes below $\mu_{\rm cut}^2=0.5$ GeV$^2$, 
though, we choose to freeze it at $\mu^2=\mu_{\rm cut}^2$. An alternate set 
of parton distribution functions used below is the 
Gl\"{u}ck, Reya and Vogt GRV98 set \cite{Gluck:1998xa} which evolves from 
$\mu_0^2=0.4\ {\rm GeV}^2$ with parametrizations provided 
above $\mu_{\rm cut}^2=0.8$ GeV$^2$. In principle, CTEQ6 covers 
the parameter space of global PDF-related data within 
conservative errors. On the other hand, the evolution with three quark flavours ($u,d,s$)
of GRV98 matches better with our evaluation 
of $\sigma_{\rm CC}({\nu_{\tau}}N)$ at low neutrino energies where we
assume light sea quarks only. In comparison, the CTEQ6 PDFs are evolved with a variable 
flavour number which is, strictly, not fully
compatible with our approach. When we quantify the uncertainties from
the PDF degrees of freedom we will, therefore, employ GRV98 for an independent comparison. 
Thereby, we can convince ourselves that the slight inconsistency 
mentioned above leads to no noticeable bias in practice. 
The CTEQ6  NLO $\overline{\rm MS}$
set of parton distribution functions is our default choice
in the evaluations below. If not stated otherwise, the curves
correspond to $\mu = Q$.

\begin{figure}
\psfig{figure=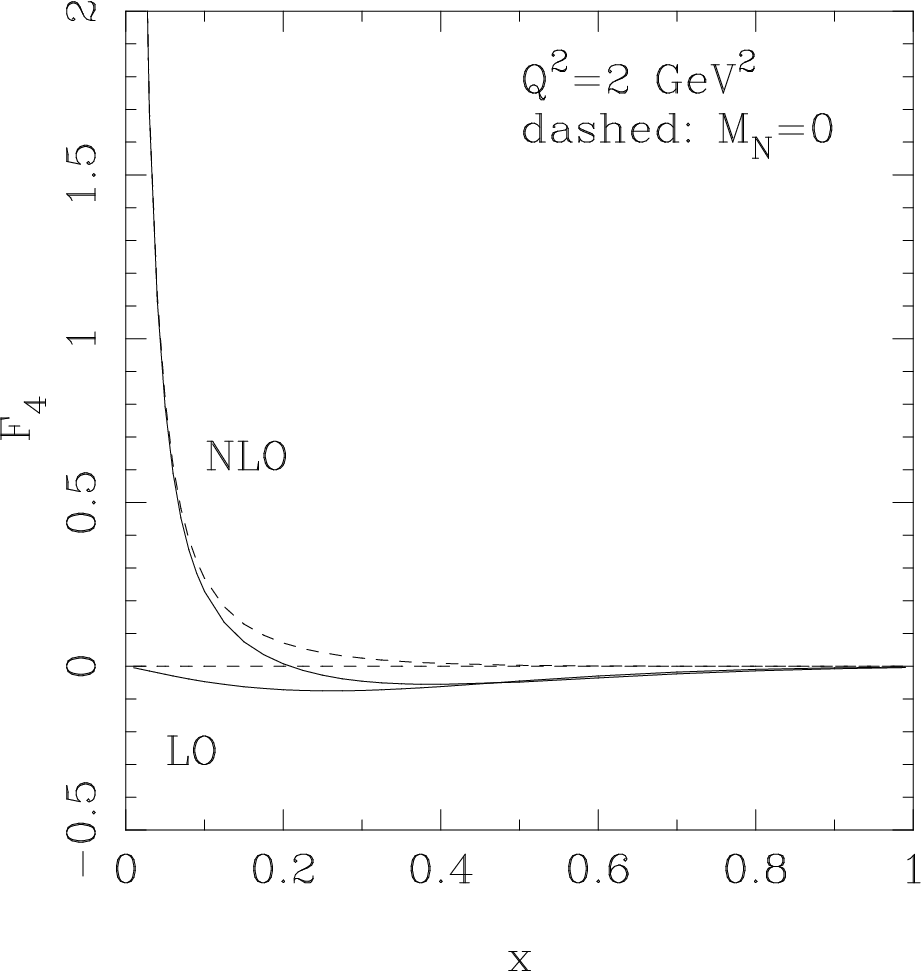,width=15cm,angle=270}
\caption{The structure function $F_4$ as a function of $x$ at fixed
$Q^2=2\ {\rm GeV}^2$, at LO and NLO, using the CTEQ6
parton distribution functions. Dashed lines show the
case with target mass $M_N=0$.
\label{fig:f4}
}
\end{figure}

In Fig.~\ref{fig:f4}, we show $F_4(x,Q^2=2\ {\rm GeV}^2)$. The solid lines in the
figure include target mass corrections
of Eq.~(\ref{eq:mix4}), 
while the dashed lines are with the 
target mass set to zero, so that $\eta\rightarrow x$ and $\rho=1$. 
The leading order (LO) curve with $M_N=0$ shows $F_4=0$, even with charm
mass corrections included. A leading order violation would occur
if the initial quark masses were set to non-zero values.  Including the
target mass corrections shows the effect of the mixing of ${\cal F}_4$ with
${\cal F}_2$ and ${\cal F}_5$ in Eq.~(\ref{eq:mix4}). The NLO corrections
have an effect primarily at small-$x$. 

\begin{figure}
\psfig{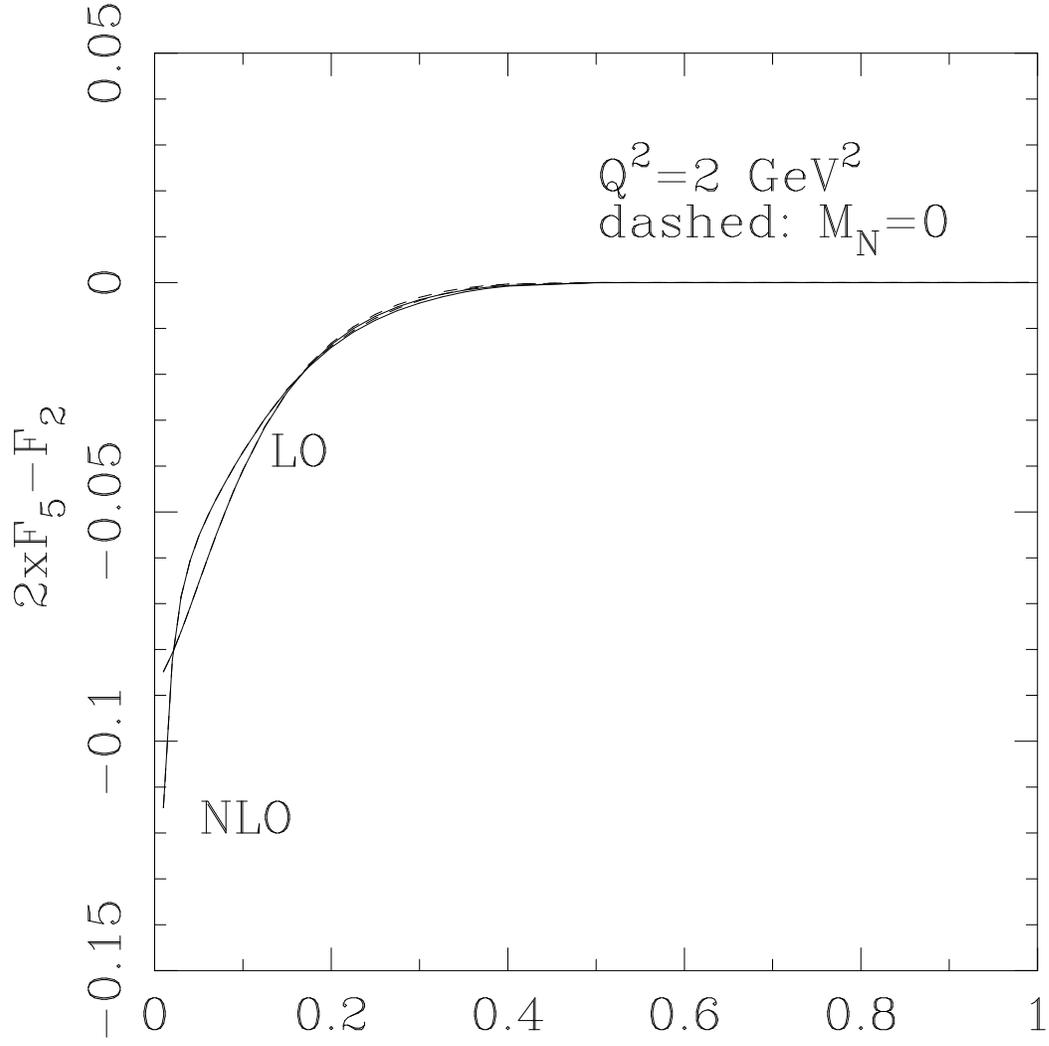}
\caption{The structure function difference $2xF_5-F_2$ as a function
of $x$ at fixed 
$Q^2=2$ GeV$^2$, at LO and NLO, using the CTEQ6
parton distribution functions. Dashed lines show the
case with target mass $M_N=0$.
\label{fig:f5}
}
\end{figure}

Fig.~\ref{fig:f5} shows the violation of the second Albright-Jarskog relation,
at fixed $Q^2=2$ GeV$^2$.
Even at leading order, $2xF_5-F_2\neq 0$. This is due to charm quark
mass corrections in $W^+ s'\rightarrow c$. The magnitude at small-$x$ reflects
the impact of the $s'$ sea distribution.
Target mass effects incorporated by Eq.~(\ref{eq:mix5})
are not significant. Including the NLO corrections makes small changes
to the curve at small-$x$. 
Both of the figures show that in evaluations of the total charged current
cross section, the naive Albright-Jarlskog relations are good approximations
to the NLO results. This is true at low energies, where
$\sigma_{\rm CC}({\nu_{\tau}}N)$ does not probe small-$x$ and at high
energies where $F_{4,5}$ are suppressed, anyway.

\subsection{Cross Sections}
\label{sec:xsec}

\begin{figure}
\epsfig{figure=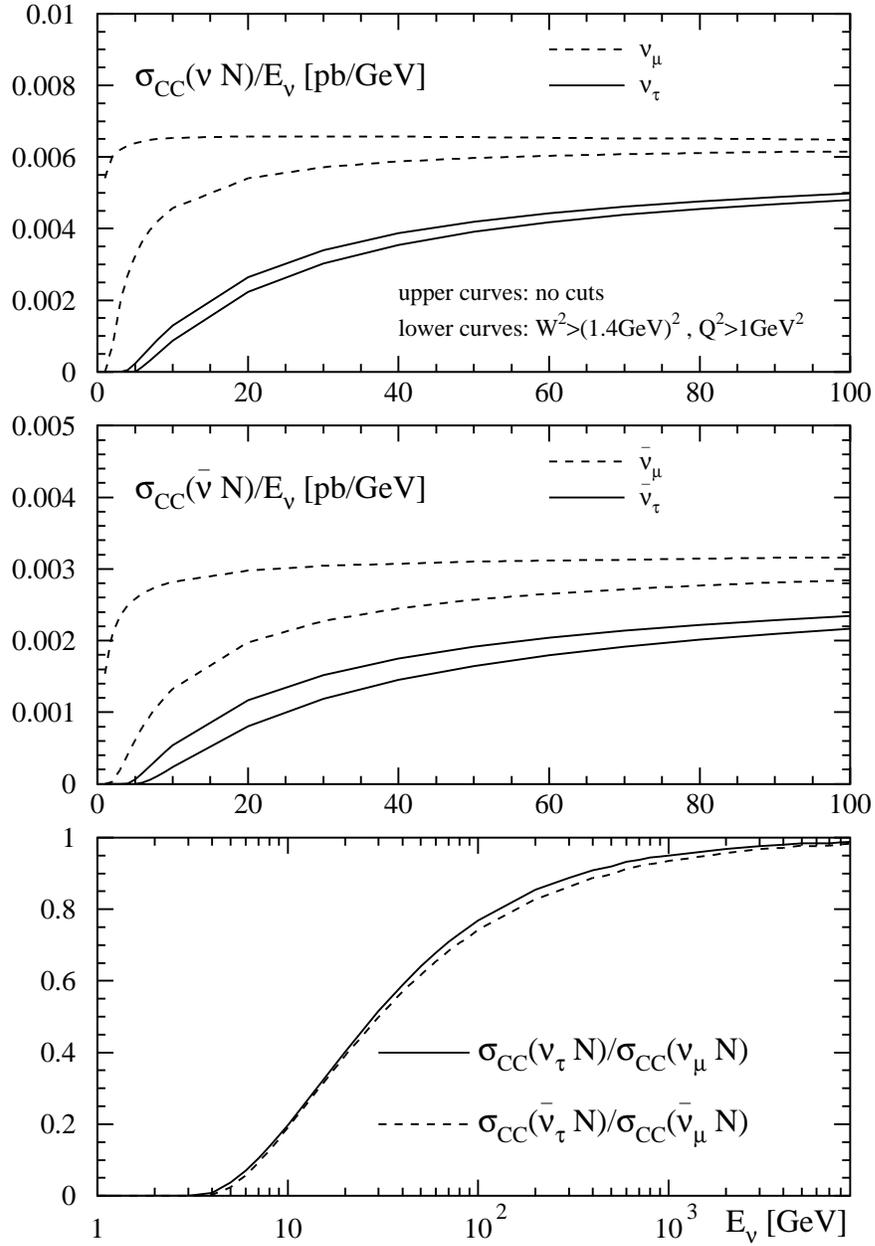,width=15cm}
\caption{Cross sections for inclusive 
neutrino [$\sigma_{\rm CC} (\nu N)$, 
upper panel] and 
anti-neutrino [$\sigma_{\rm CC} ({\bar \nu} N)$, middle panel]
production of charged leptons
on an isoscalar target, evaluated at NLO and plotted
versus the energy of incident $\mu$- and
$\tau$-flavoured (anti-)neutrinos. We show the outcome of a naive 
integration over the full kinematic range of $W^2$ and $Q^2$ along 
with the effect of imposing DIS cuts on these variables. 
The ratios 
$\sigma_{\rm CC} (\nu_{\tau} N ) / \sigma_{\rm CC} (\nu_{\mu} N )$
and 
$\sigma_{\rm CC} ({\bar \nu}_{\tau} N ) / \sigma_{\rm CC} ({\bar \nu}_{\mu} N )$
which we show in the lower panel are insensitive to these cuts 
(and we show the uncut
results).
\label{fig:sig}
}
\end{figure}

\begin{figure}
\epsfig{figure=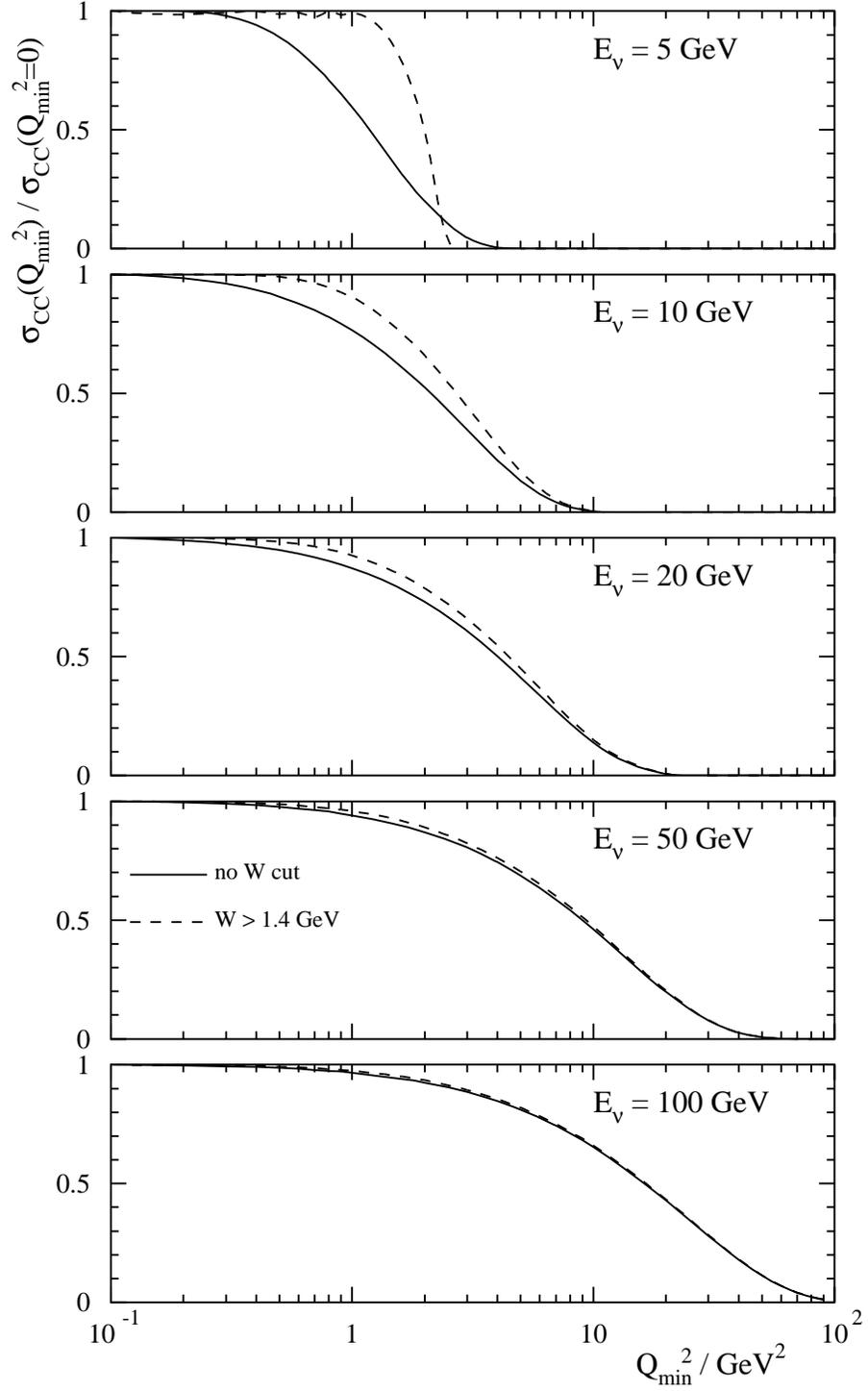,width=15cm}
\caption{Ratio $\sigma_{\rm CC}(Q^2_{\min})/\sigma_{\rm CC}(Q^2_{\min}=0)$ versus $Q^2_{\min}$
for $\nu_\tau N$ CC interactions with $E_\nu=5,$ 10, 20, 50 and 100 GeV. 
The solid lines are with $W_{\min}=0$, the dashed lines with $W_{\min}=1.4$ 
GeV.
\label{fig:qmin}
}
\end{figure}

The cross sections for neutrino and antineutrino charged current interactions
with isoscalar nucleons are shown in the first two panels of 
Fig.~\ref{fig:sig}. We
use our default set of CTEQ6 parton distribution functions with the
factorization scale equal to the renormalization scale $\mu^2$,
and $\mu^2=Q^2$. Below $\mu^2=\mu_{\rm cut}^2$, we set 
$\mu^2=\mu_{\rm cut}^2$ in the parton 
distribution functions and in $\alpha_s$, but we keep the explicit $Q^2$ 
dependence in the differential cross section of Eq.~(\ref{eq:nusig}) 
as well as in the counter-logs $\sim \ln Q^2 / \mu_{\rm cut}^2$
of the NLO coefficient functions. For $Q^2 < \mu_{\rm cut}^2$, a noticable 
impact of these logs is indicative of long-distance
strong interaction.
The evaluation can, therefore, not be trusted perturbatively,
whenever it becomes sensitive to the technical choice $\mu > \mu_{\rm cut}$.
Most of our results below are, however, completely insensitive to it.
In Fig.~\ref{fig:sig}
muon (anti-)neutrino cross sections appear with dashed
lines, while the solid lines show the tau (anti-)neutrino cross sections.
The upper curves show no cuts while the lower curves have 
$W^2 =Q^2 (x^{-1}+1) +M_N^2>(1.4\ {\rm GeV})^2$ and $Q^2> 1$ GeV$^2$.
As we show in Fig.~\ref{fig:qmin}, for $W_{\min}=1.4$
GeV, the tau neutrino CC cross section is fairly insensitive to the $Q^2$ cut
of 1 GeV$^2$. At, for example, $E_\nu=20$ GeV, $\sigma_{\rm CC}(Q_{\min}^2=1\ {\rm GeV}
^2)/\sigma_{\rm CC}(Q_{\min}^2=0)=0.93$ for $\nu_\tau N$ CC interactions.
The $Q^2$ cut has a larger impact on $\sigma_{\rm CC}(\nu_\mu N)$, 
where
$\sigma_{\rm CC}(Q_{\min}^2=1\ {\rm GeV}
^2)/\sigma_{\rm CC}(Q_{\min}^2=0)=0.85$ for $E_\nu=20$ GeV, due to the fact
that for nearly massless leptons, $Q^2>Q^2_{\min}$
cuts out a larger fraction of the available phase space. 

The small changes in the CC cross sections
with $Q_{\min}^2= 1\ {\rm GeV}^2$ lead one to expect that non-perturbative effects
at low $Q^2$ are unlikely to be large when one also applies the
$W_{\min}$ cut of 1.4 GeV. At low energies, without the $W_{\min}$ cut,
a substantial contribution to the cross sections comes from $Q^2< 1$ GeV$^2$,
however, and this is precisely where the DIS cross section is only a rough
approximation to the true cross sections with quasi-elastic and resonant
as well as non-resonant contributions.

The bottom panel of Fig.~\ref{fig:sig} shows the ratio of the $\nu_\tau N$ to 
$\nu_\mu N$ CC cross sections (solid lines) and the same ratio for
antineutrinos. Shown are the uncut results, but the results with the
$W_{\min}$ and $Q^2$ cuts agree to within 3\% for
$E_\nu>20$ GeV. Of note is the
fact that even at $E_\nu=10^3$ GeV, the $\nu_\tau N$ to 
$\nu_\mu N$ CC cross section ratio is still 5\% below
unity. At 100 GeV, the ratio is 0.76. There are two reasons for
the deficit in the $\nu_\tau$ CC cross section: the reduced phase
space and the contribution of $F_5$. 
The reduced phase space is reflected in the integration limits for
$x$ and $y$ (Eqs.~(\ref{eq:limitx}) and (\ref{eq:limity})) \cite{Starkov:sa}.
This is
responsible for about half of the suppression of the $\nu_\tau$ 
CC cross section relative to $\sigma_{\rm CC}(\nu_\mu N)$. In Eq.~(\ref{eq:nusig}),
the $F_5$ term appears with a minus sign, and no factor of $x$.
Since $F_5\sim F_1\sim q(x,Q^2)$, there is a small-$x$ enhancement
of its contribution to the cross section at high energies. The $F_5$
term accounts for the rest of the suppression of the $\nu_\tau$ cross
section at high energies.
The tau mass corrections to the
prefactors of $F_1,\ F_2$ and $F_3$ become
negligible at high energies because the 
low-$x$ rise of $q(x)$
is tempered by factors of $x$ or $y$ for these structure function.

Since mass effects for tau neutrino interactions persist to 1 TeV,
it is interesting to compare to the case of muon neutrino CC interactions
at low energies where the muon mass is important. Since the factor
of lepton mass comes into the equations via $m_\tau^2/E_\nu$, the
energy for muon neutrino interactions equivalent to 1 TeV for $\nu_\tau$
is $E_\nu= 3.5$ GeV. At this energy, $\sigma_{\rm CC}(\nu_\mu N)$ including
$m_\mu=0.106$ GeV (without cuts) is 2\% lower than the cross section
with $m_\mu=0$. This smaller suppression is due to the fact that
at low energies, one does not get significant small-$x$ contributions
to $F_5$. Furthermore, for $\nu_\mu N$ scattering, this energy is
in the range where the DIS approximation to the total cross section
is not reliable.

\begin{figure}
\psfig{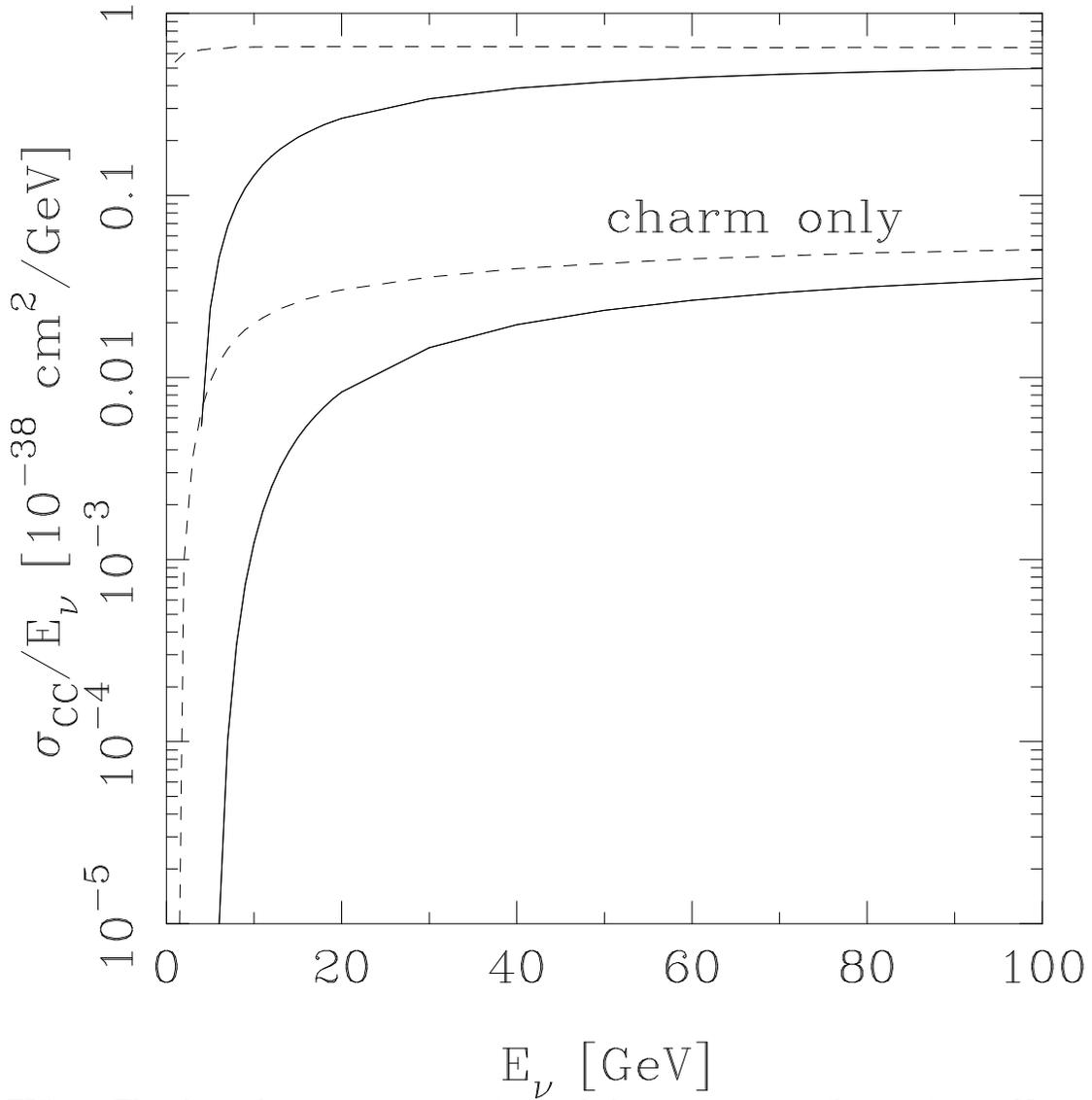}
\caption{The charged current cross section and the separate contribution
from $\nu N\rightarrow c X$ for incident $\nu_\mu$ (dashed line)
and $\nu_\tau$ (solid line). At
$E_\nu=100$ GeV, the charm production contribution is about 7\% of the
total.
\label{fig:charm}
}
\end{figure}

Charm production in neutrino interactions is a small contribution.
In Fig.~\ref{fig:charm}, we show the total charged current cross section and
the separate cross section for $\nu N\rightarrow c X$ as a function of
incident neutrino energy. The dashed curves are for incident $\nu_\mu$,
the solid curves for incident $\nu_\tau$. At $E_\nu=100$ GeV,
charm production contributes about 7\% of the cross section for both
$\nu_\mu$ and $\nu_\tau$.

\begin{figure}
\epsfig{figure=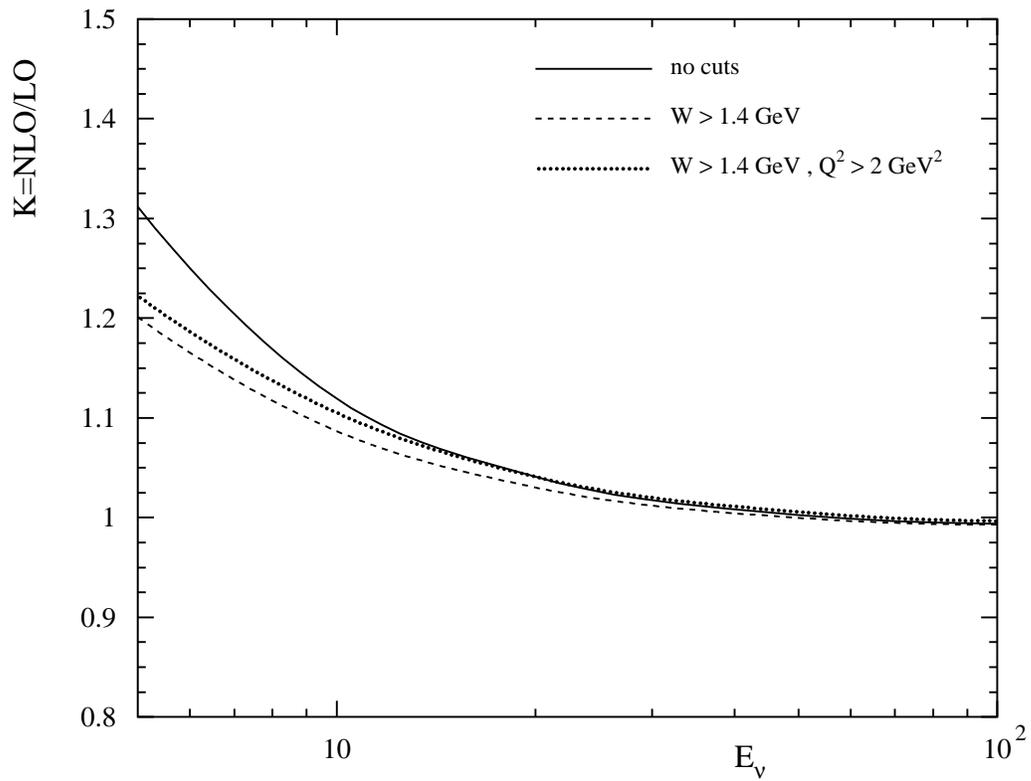,width=15cm}
\caption{The K-factor $K=$NLO/LO for tau neutrinos with no cuts
(solid), $W_{\min}=1.4$ GeV (long dash) and $W_{\min}=1.4$ GeV with
$Q^2>2$ GeV$^2$ (short dash).
\label{fig:kfac}
}
\end{figure}

The K-factor, a comparison of the NLO to LO\footnote{The LO result is
obtained by neglecting all \oalone\ terms in Section \ref{sec:formulae} and
employing the CTEQ6L PDFs.} charged
current cross sections for incident tau neutrinos is shown in Fig.~\ref{fig:kfac}.
As to be expected, NLO corrections are most significant near threshold
where virtualities $Q^2$ are lower and $\alpha_s$ is larger than at higher
energies. At $E_\nu=10$
GeV, $K=$NLO/LO=1.12, reducing to $K=1$ at $E\simeq 50$ GeV for
the evaluation with no cuts.
The three curves show that with $W_{\min}>1.4$ GeV and $Q^2>2$ GeV$^2$,
the K-factor is the same as for the uncut cross section except
for $E_\nu <10$ GeV where the cuts improve the perturbative convergence. 
The K-factor with only the $W_{\min}$ cut is
even slightly lower, confirming  the expectation based on  
Fig.~\ref{fig:qmin} that perturbation theory 
is well behaved and its convergence
does not have to be improved by
any further $Q_{\min}$ cut on top of $W_{\min}$. Overall, the $K$ factor 
is reasonably close to one to indicate trustworthy perturbative
NLO predictions, even more so when the $W_{\min}$ cut is
imposed.

\subsection{Uncertainties}
\label{sec:errors}

In this Section we investigate a few factors that cause
theoretical errors in the evaluation of $\sigma_{\rm CC}$.
A full assessment of uncertainties will have to be based on 
the combination of the quasi-elastic and resonant channels with our 
DIS results. At higher energies, though, where DIS 
becomes dominant, we already can provide a good guide
towards error estimates.

\begin{figure}
\epsfig{figure=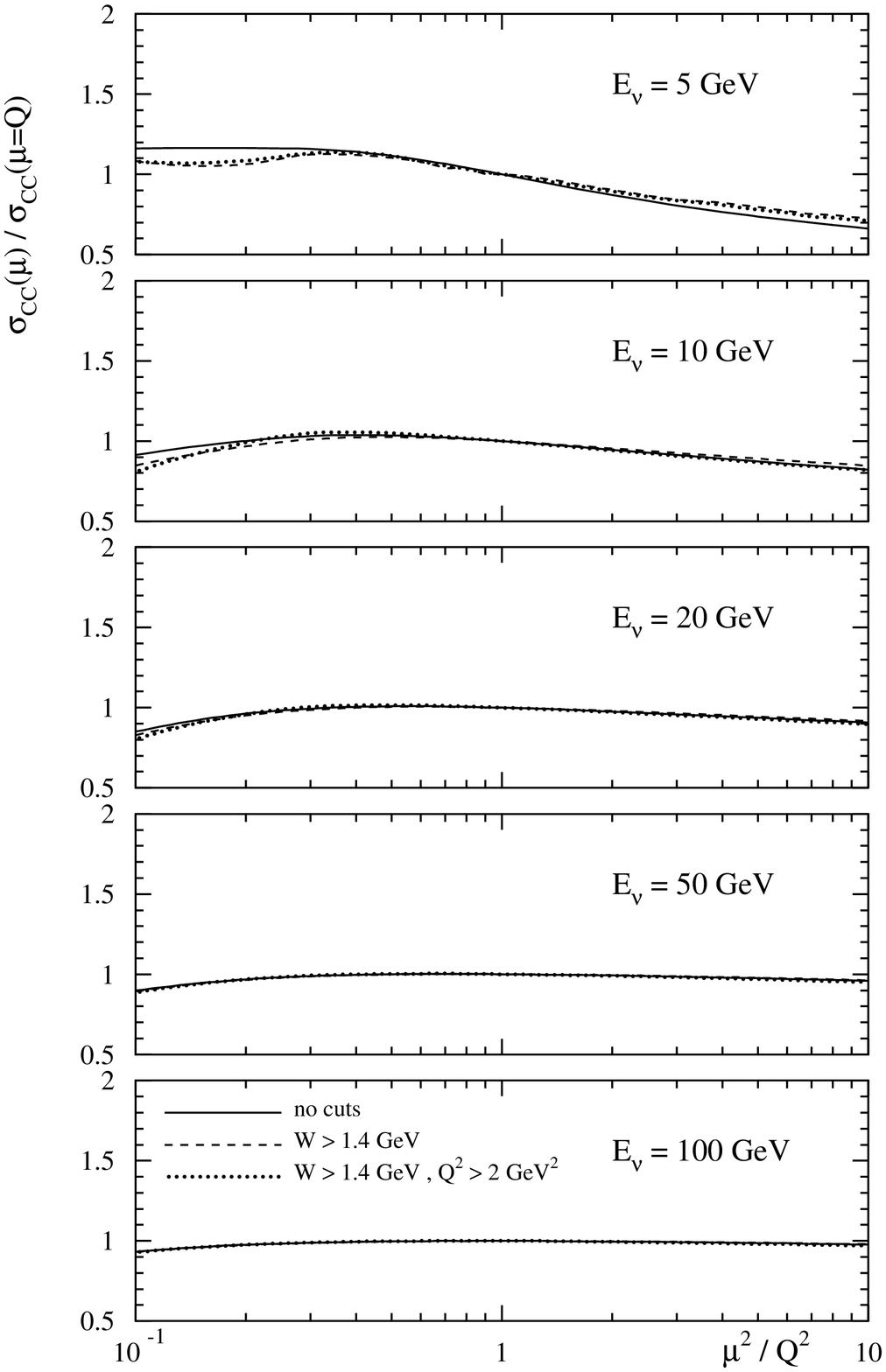,width=15cm}
\caption{The ratio of $\sigma_{\rm CC}(\mu)/\sigma_{\rm CC}(\mu=Q)$ in NLO for $\nu_\tau$N
interactions, as a function of $\mu^2/Q^2$ for several values of $E_\nu$.
\label{fig:mu}
}
\end{figure}

\begin{figure}
\epsfig{figure=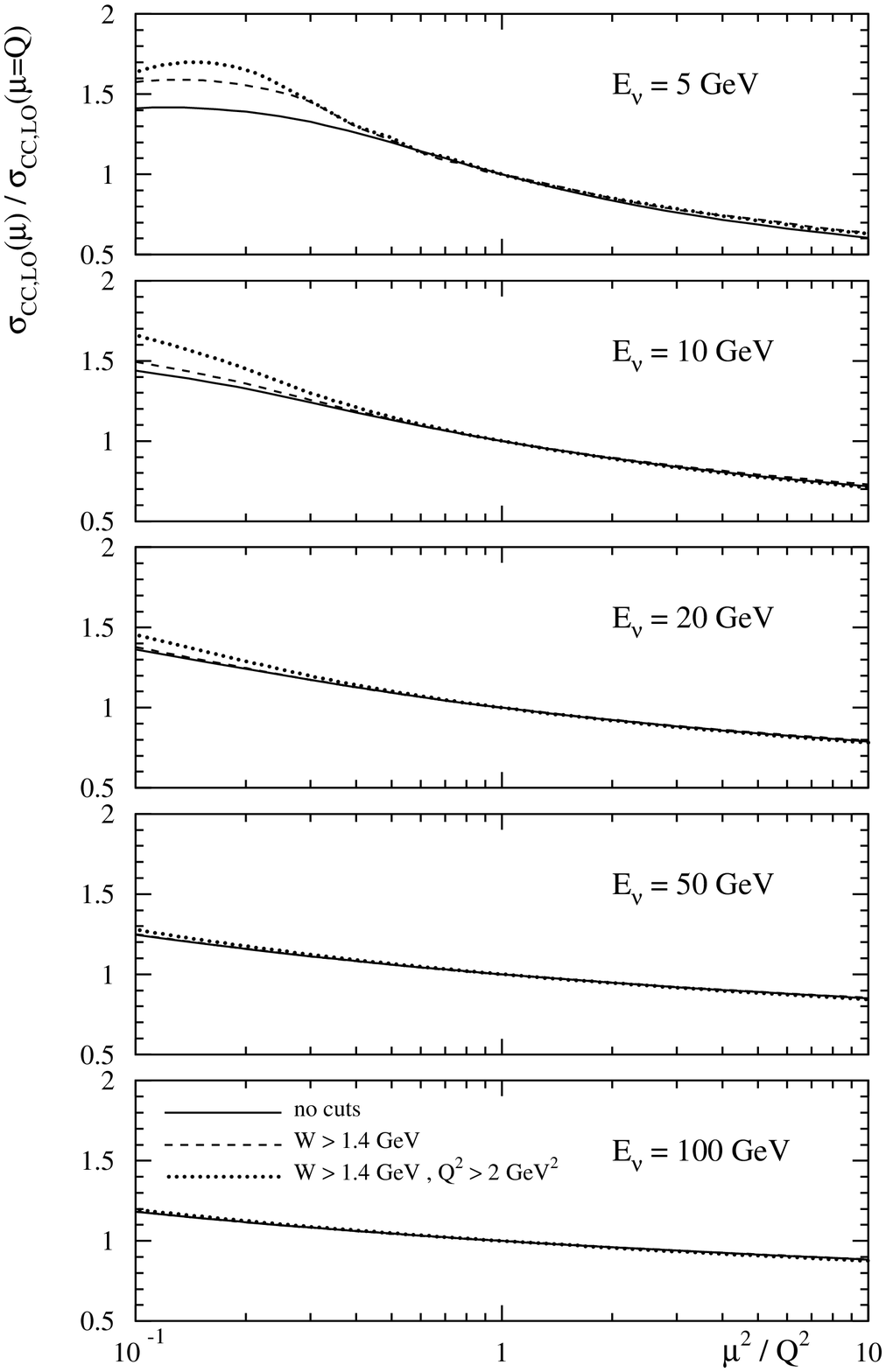,width=15cm}
\caption{The ratio of $\sigma_{\rm CC}(\mu)/\sigma_{\rm CC}(\mu=Q)$ in LO for $\nu_\tau$N
interactions, as a function of $\mu^2/Q^2$ for several values of $E_\nu$.
\label{fig:mulo}
}
\end{figure}

\begin{figure}
\epsfig{figure=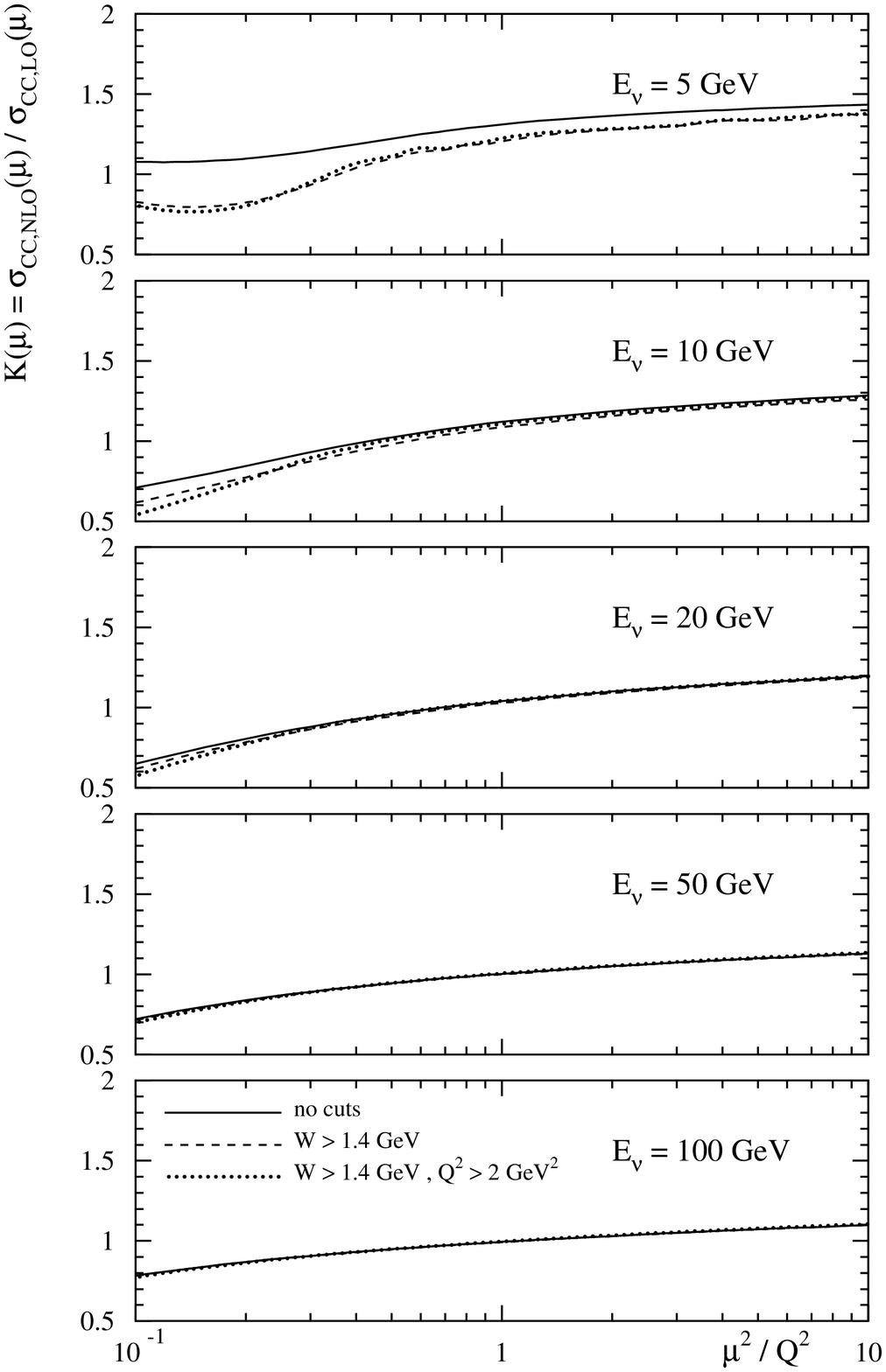,width=15cm}
\caption{The K-factor $K=$NLO/LO versus factorization 
scale $\mu$ for tau neutrinos with no cuts
(solid), $W_{\min}=1.4$ GeV (long dash) and $W_{\min}=1.4$ GeV with
$Q^2>2$ GeV$^2$ (short dash).
\label{fig:kfac2}
}
\end{figure}

An uncertainty in our evaluation of the
cross section is due to  the factorization scale dependence 
of the cross section at fixed order in perturbation theory. 
To evaluate the uncertainty due to the scale dependence, we have
varied the factorization and renormalization scale ($\mu=\mu_F=\mu_R$) 
over a very wide range of
$\mu^2=0.1-10\ Q^2$ for energies between 5 GeV and 100 GeV.
A selection of energies is shown in Fig.~\ref{fig:mu}. We show the ratio of 
$\sigma_{\rm CC}(\nu_\tau N)$ as a function of
$\mu^2/Q^2$ to the cross section
at $\mu^2=Q^2$. The flat $\mu$ dependence is a reassuring feature
of the NLO calculation as opposed to the monotonic decrease
with $\mu$ observed in LO, see Fig.~\ref{fig:mulo}.
While we discuss here the full range of $\mu^2$ as plotted 
in Fig.~\ref{fig:mu}, the perturbative
stability observed in Fig.~\ref{fig:kfac}
suggests that one would very likely overestimate
the uncertainty from higher orders by such a wide scale variation. 
This seems even more so if we look at the scale dependent 
K-factor in Fig.~\ref{fig:kfac2} which prefers 
scales close to the canonical DIS choice $\mu=Q$ where $K\simeq 1$. 
Still, scale choices are arbitrary to a large extent
and we prefer to discuss the full picture
instead of narrowing it down to some window for $\mu$ around $\mu=Q$.
At $E_\nu=5$ GeV, the variation is the largest for the
cross section evaluated with no cuts, ranging between 1.16 and 0.66.
As explained in Section \ref{sec:sf},
for $\mu^2<\mu_{\rm cut}^2$, we always set $\mu=\mu_{\rm cut}$\footnote{
{\it I.e.}~strictly the plot should be labeled $\sigma_{\rm CC} \left( \max \{ \mu , \mu_{\rm cut}\}
\right) / \sigma_{\rm CC} \left( \max \{ Q , \mu_{\rm cut}\}
\right)$. 
}. 
For small $\mu/Q$, the ratio without cuts in $\{W,Q\}$ is nearly
constant as a function of $\mu^2/Q^2$, 
showing the degree to which the uncut cross section comes from
small $\mu$ values where $\mu_{\rm cut}$ takes over and where 
the perturbative treatment is unreliable.  
With $\{W,Q\}$ cuts applied, the variation over the range of
scales is slightly less. The
scale uncertainty at low energies
underlies a larger uncertainty associated with the DIS approximation near
threshold where the quasi-elastic and resonant contributions are significant.

At higher energies, the variation in the ratio is between $\sim 0.85-1$.
For $E_\nu=20$ GeV, the ratio in Fig.~\ref{fig:mu} is between 0.95 and 1 for
$\mu^2/Q^2\sim 0.2-4$.

The uncertainty in the cross section due to uncertainties in the 
parametrization of the parton distribution functions (PDF)
is harder to quantify than the scale dependence. 
One approach is to use different
parton distribution functions than the default CTEQ6 set.
As a comparison, the GRV98 PDF set, at $E_\nu=20$ GeV
with $\mu=Q$,
yields a cross section only about one percent smaller than the CTEQ6
set. At $E_\nu=5$ and 10 GeV, the GRV98 cross sections are 10\% and
4\% lower than the CTEQ6 cross sections when no cuts are applied. The larger
deviation at lower energies is due to the different high-$x$ distributions
in the two PDF sets. Including $W>1.4$ GeV results in a deviation
of 3.6\% at $E_\nu=5$ GeV, even less for higher energies.
Table \ref{tab:errors} represents results for a few selected energies
with a cut in $W$ applied.
As commented above in Section \ref{sec:sf},
the PDFs  also differ in their treatment of the number of active flavors
and in the value of the strong coupling constant.

The CTEQ collaboration has also provided distributions in addition to
their best fit set \cite{Pumplin:2002vw}. 
The 20 dimensional parameter space to which the PDFs
are sensitive yields 40 PDF sets with plus/minus variations on the
eigenvector directions in that 
space. The resulting error estimate
on $\sigma_{\rm CC}(\nu_\tau N)$ from evaluating the 40 sets is 3\% at
$E_\nu=20$ GeV, see Table \ref{tab:errors} for other values of $E_\nu$.
Overall, the GRV98 results lie within the uncertainty
estimate  suggested by the CTEQ6 eigenvector PDFs. We can, therefore, be
confident of the absence of a systematic effect from the number of flavours
and that the statistical uncertainties as encoded in the CTEQ6 sets 
provide a realistic PDF error estimate for $\sigma_{\rm CC}({\nu_{\tau}}N)$.
 
We have incorporated kinematic corrections 
due to including the target hadron mass $M_N$ by employing the parton
light cone fraction $\xi$, which equals the Nachtmann variable $\eta$
\cite{nacht}
at leading order in the massless quark limit. 
One finds that $\eta$ is much different than Bjorken-$x$ at large $x$.
For example, for $Q^2=2\,(10)$ GeV$^2$, $\eta=0.45\,(0.49)$ at $x=0.5$
and $\eta=0.75\,(0.92)$ at $x=1$.
The use of $\eta$ rather than $x$ in the structure functions
has the largest impact at high $x$ and low $Q^2$.

Target mass corrections are also included 
via Eqs.~(\ref{eq:mix1})-(\ref{eq:mix5}), in which the ${\cal F}_i$
are mixed with 
target mass dependent prefactors for a given $F_i$.
These formulae are based on the assignment of $p_+=\xi P_+$ to the light-cone
momentum of the massless incident parton ($p^2=0$) given $P^2=M_N^2$. 
The parton and nucleon are
assumed to have collinear momenta, $p_\perp = 0$. 
The formalism is discussed in detail in
Ref.~\cite{aot}, however, other choices for the model of including
target mass effects are possible,
for example, including parton
transverse momentum \cite{Ellis:1982cd} $p_\perp \neq 0$. 
The model of Ref.~\cite{Ellis:1982cd}
reproduces the kinematic corrections for the leading twist operator
product expansion result \cite{Georgi:1976ve}. 
As a consequence,
our results here serve only as a guide to the magnitude of target mass
corrections to the charged current cross sections.
For the transverse structure function $F_T$ of Ref. \cite{Ellis:1982cd},
which equals ${\cal F}_2$ in the collinear parton approximation,
the difference between the $p_\perp\neq 0$ approximation giving $F_T$
and our order $p_\perp^0$ approximation is less than 10\% over
all $x$ and $Q^2$ as low as 1 GeV$^2$, rapidly falling below 5\% at
$Q^2=$2 GeV$^2$.
Low energy ${\cal O}(p_\perp^2)$ effects and the target mass
treatment in general border onto dynamical higher twists
at large-$x$ \cite{rgp,Johnson:1979ty,pr}. Those concepts will be 
investigated in more detail when future work will 
combine the DIS cross-sections with the non-DIS channels.

The effects of the target mass corrections as implemented here on
$F_4$ and $2xF_5-F_2$, as shown in Figs.~\ref{fig:f4} and 
\ref{fig:f5}, are small. In terms of the total cross section, the lowest energies
are most affected. In the absence of kinematic cuts, 
$\sigma_{\rm CC}(\nu_\tau N)(E_\nu=10\ {\rm GeV})$ 
for $M_N=0$ is 8\% larger than the cross section
including the target mass via $\eta$ and Eqs.~(\ref{eq:mix1})-(\ref{eq:mix5}). For $E>20$ GeV,
the deviation is less than 5\%, down to the 2\% level at 50 GeV incident
neutrino energy.
When one includes $W>W_{\min}=1.4$ GeV, the effect of target mass
corrections on the $\nu_\tau N$ CC cross section is less than
2\% for $E_\nu>8$ GeV. The reduced effect is due to the fact that the
$W_{\min}$ value reduces the region of integration for large $x$
since $2M_N E_\nu y(1-x)\geq W_{\min}^2-M_N^2$.

\section{Conclusions}
\label{sec:sum}

\begin{figure}
\epsfig{figure=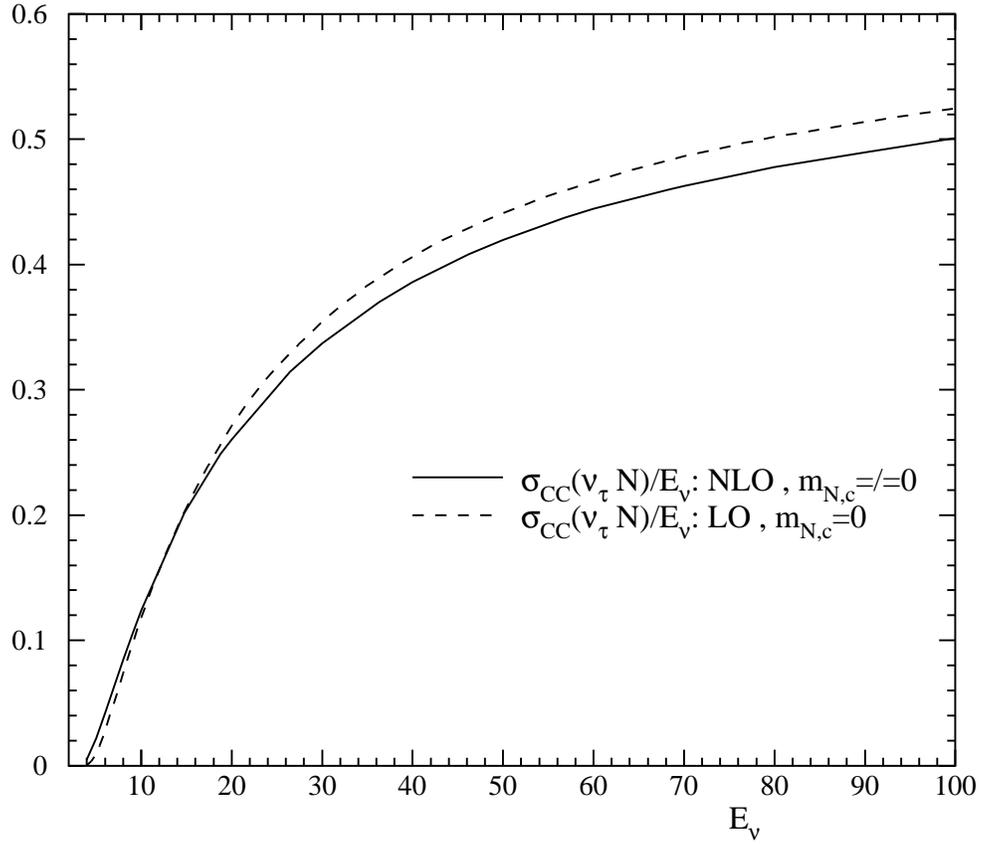,width=15cm}
\caption{Our calculation of $\sigma_{\rm CC}({\nu_\tau} N)/E_\nu$  
versus neutrino energy compared to its naive evaluation
neglecting masses and NLO corrections.
\label{fig:naive}
}
\end{figure}

The NLO corrections for the $\nu_\tau N$ charged current cross section
have a relatively large impact at low neutrino energies near threshold
and less of an impact at high energies. 
We have already shown the K-factor in Fig.~\ref{fig:kfac}. 
In Fig.~\ref{fig:naive}, we show $\sigma_{\rm CC}
(\nu_\tau N)/E_\nu$
versus neutrino energy at NLO including target mass 
corrections as implemented via Eqs.~(\ref{eq:mix1})-(\ref{eq:mix5})
and charm mass correction.
This is compared to the naive evaluation, neglecting masses and NLO 
corrections, where the Albright-Jarlskog relations are correct.
At high energies, the target mass corrections are negligible. The main effect
is due to the charm mass threshold. At lower energies, the larger QCD
K-factor is compensated by the reduction in the cross section due to
target mass and charm mass effects.

In the theoretical evaluation of the $\nu_\tau N$ charged current
cross section, the scale dependence in the PDFs and $\alpha_s$ is
a large uncertainty at low energies. The parametrization of the PDFs
does not seem to be a large uncertainty in the evaluation of the total
cross section, especially when one applies DIS cuts on $W^2$ and $Q^2$.

Target mass corrections and the importance of $Q^2<1$ GeV$^2$ are small
at high energies, but they are significant in the 10-20 GeV energy range
and lower, especially for the cross section without kinematic cuts.
Our implementation of target mass corrections neglects parton
transverse momentum ($p_\perp^2$). This approach makes the inclusion
of NLO QCD corrections straightforward, however, it neglects
corrections of order $M_N^2/Q^2$ induced by nonzero $p_\perp^2$ 
\cite{Ellis:1982cd}. 
As a consequence, one should view our target
mass corrections as approximate. 

In any case, the 
DIS cross section in the energy region below
$E_\nu\sim 10-20$ GeV is difficult to interpret. At best, the
uncut cross section is a
crude approximation to the total cross section including resonant
hadron production, {\it e.g.}, $\Delta$ production.
The cross section with cuts is likely a better representation of the
non-resonant neutrino-nucleon interactions, however, the issue of
avoiding double-counting in combining resonant and non-resonant interactions
is not solved theoretically. Phenomenological approaches are being
explored \cite{by} for applications to $\nu_\mu N$ interactions
in the few GeV region that may also be applied to $\nu_\tau N$ interactions
at higher energies. The universality of $M_N^2/Q^2$ corrections, carried 
over from copious electromagnetic interaction data to the
context of weak interactions,
is not completely clear. 

The $\nu N$ cross section is an important ingredient in current and future
atmospheric and neutrino factory experiments. Our evaluation of the
NLO corrections for $\nu_\tau$N CC interactions including charm mass
corrections and an estimate of target mass effects is part of a larger
theoretical program to understand the inelastic $\nu N$ cross section
over the full energy range relevant to experiments.

\acknowledgments
We thank K.~Ellis, M.~Goodman, R.~Jakob, F.~Olness, W.K.~Tung  
and U.K.~Yang for discussions.
This research was supported in part by the
National Science Foundation under Grants 
PHY-0070443 and PHY-9802403 and DOE Contract No. FG02-91ER40664.

\newpage
\setcounter{equation}{0}
\def\theequation{A\arabic{equation}}
\section*{Appendix A: The Albright-Jarlskog Relations}
\label{sec:app}

In this Appendix we pinpoint the approximations for which 
the Albright-Jarlskog relations (AJRs) (\ref{eq:aj1}) and (\ref{eq:aj2}) 
hold beyond the naive parton model (where they are exact). 
Below, we will rewrite the AJRs in terms of helicity amplitudes.
First, however, we can immediately tell from Eqs.~(\ref{eq:mix1})--(\ref{eq:mix5}) that
the mixing of $\{ {\cal F}_2^c, {\cal F}_4^c, {\cal F}_5^c \}$
for $M_N \neq 0$ will violate both (\ref{eq:aj1}) and (\ref{eq:aj2}).
We will, therefore, restrict the following discussion to $M_N=0$.
The charm mass will be retained to trace its (somewhat less obvious) 
impact on the AJRs.
In terms of helicity projections 
\begin{eqnarray}
W_0 &  = &\varepsilon_0^{\mu}\varepsilon_0^{\nu}W_{\mu\nu}\\
W_s &  =& \Bigl(\varepsilon_0^{\mu}\varepsilon_q^{\nu} +
\varepsilon_q^{\mu}\varepsilon_0^{\nu}\Bigr)W_{\mu\nu}\\
W_q & = &\varepsilon_q^{\mu}\varepsilon_q^{\nu}W_{\mu\nu}\\
W_{\pm} & = &\varepsilon_{\pm}^{\mu}{\varepsilon_{\pm}^{\nu}}^{\ast} W_{\mu\nu}
\end{eqnarray}
with polarization vectors in terms of virtual W momentum $q$ and
longitudinal reference vector $k$
\begin{eqnarray}
\varepsilon_{\pm}^\mu & = \frac{1}{\sqrt{2}}\ (0,\mp 1,-i,0) &\ {\rm \it (transverse)} \\
\varepsilon_q^\mu &  ={q^\mu\over \sqrt{-q^2}} &\ {\rm \it (scalar)} \\
\varepsilon_0^\mu &  
={(-q^2) k^\mu+(k\cdot q) q^\mu\over \sqrt{(-q^2)[(k\cdot q)^2]}}
&\ {\rm \it (longitudinal)} 
\end{eqnarray}
the tensor basis ${\cal F}_{i=1,...,5}^c$ can be written as:
\begin{eqnarray}
{\cal{F}}_1^c & & = {1\over 2}(W_+ + W_-) \\
{\cal{F}}_2^c & & = {\lambda \over 2}(W_+ + W_- + 2 W_0) \\
{\cal{F}}_3^c & & = \mp {1\over 2}(W_+ - W_-) \\
{\cal{F}}_4^c & & = {1\over 2}(W_0 + W_q - W_s) \\
{\cal{F}}_5^c & & = {1\over 2}(W_+ + W_- + 2 W_0 - W_s) \ .
\end{eqnarray}
We then see that
\begin{equation}
F_2^c - 2 x F_5^c = 2 \frac{x}{\lambda} {\cal{F}}_2^c - 2 x {\cal{F}}_5^c
=  x W_s
\end{equation}
singles out the interference term $W_s$ between scalar 
and longitudinal polarization.
The latter involves 
a contraction with $\varepsilon_q^{\mu} \propto q^{\mu}$. Now, as 
long as we make the single $W$ boson exchange approximation, the 
DIS process
is equivalent to an effectively abelian (electroweak) interaction {\it if}
all quarks are massless (or of the same mass). 
Then, $F_2 - 2 x F_5=0$ is guaranteed by naive
gauge invariance under $\varepsilon^{\mu}_q \to \varepsilon^{\mu}_q - q^{\mu} / \sqrt{-q^2}
= 0^{\mu}$. This is a stronger statement than helicity conservation
for massless spin-1/2 quarks because gauge
invariance holds to any order in $\alpha_s$ while helicity conservation
breaks down when non-collinear
NLO radiation generates angular momentum.
A transition $W^+ s \rightarrow c$ with $0=m_s \neq m_c$, however,
is necessarily non-diagonal in flavour space 
and the interaction cannot be abelianized. Thus, naive gauge invariance does not hold
and the second AJR (\ref{eq:aj2}) is not protected from charm mass corrections. It will,
however, hold at any order in $\alpha_s$ for massless quarks.

Worth mentioning is also that 
$F_4^c = 0 $ at \oalzero\ 
even for the charm production process, indicating that
$W_0+W_q-W_s=0$ at LO. This should be compared to the longitudinal structure function
$F_2 - 2x F_1\propto W_0$ which does not vanish by helicity conservation
in LO when a massive charm quark is produced in the final state.  
Now, $F_4$ is obtained from
\begin{equation}
W_0 + W_q - W_s={k^\mu k^\nu\over Q^2} W_{\mu\nu} \ .
\end{equation}
At the parton level, $k$ is the {\it incoming} parton momentum
$p$. $F_4$ does, therefore, not receive corrections from the final 
state charm mass in LO as long as initial state down 
and strange masses are zero ($p^2=0$).

\newpage

%
\begin{table}[h]
\begin{tabular}{lllll}
$i$ & $A_i                           $ & $B_{1,i}               $ & $B_{2,i}            $ & $B_{3,i}        $ \\  \hline 
$1$ & $0                             $ & $1-4\xi+\xi^2          $ & $\xi-\xi^2          $ & $\frac{1}{2}    $ \\
$2$ & $K_A                           $ & $2-2\xi^2-\frac{2}{\xi}$ & $\frac{2}{\xi}-1-\xi$ & $\frac{1}{2}    $ \\
$3$ & $0                             $ & $-1-\xi^2              $ & $1-\xi              $ & $\frac{1}{2}    $ \\ 
$5$ & $\frac{\lambda -1-K_A}{\lambda}$ & $-1-\xi^2              $ & $3-2\xi-\xi^2       $ & $\xi-\frac{1}{2}$ \\  
\end{tabular}
\caption{\label{thq}Coefficients for the expansion of $h_i^q$ in (\ref{eq:smallhq})}
\end{table}

\begin{table}[h]
\begin{tabular}{lllll}
$i$ & $C_{1,i}$ & $C_{2,i}$ & $C_{3,i}$ & $C_{4,i}$ \\  \hline
$1$ & $4-4(1-\lambda)$ & $\frac{(1-\lambda )\xi}{1-\lambda \xi}-1$ & $2$ & $-4$ \\
$2$ & ${8-18(1-\lambda ) \atop +12(1-\lambda )^2}$ & $\frac{1-\lambda}
{1-\lambda \xi}-1$ & $6\lambda$ & $-12 \lambda$ \\
$3$ & $2(1-\lambda )$ & $0$ & $-2(1-\xi)$ & $2$ \\ 
$5$ & $8-10(1-\lambda )$ & $\frac{(1-\lambda )\xi}{1-\lambda \xi}-1$ &$4$& $-10$ \\
\end{tabular}
\caption{\label{thg}
Coefficients for the expansion of $h_i^g$ in (\ref{hig})}
\end{table}

\begin{table}[h]
\begin{tabular}{ccc}
$E_{\nu}\ [{\rm GeV}]$ & $\Delta\ \sigma_{\rm CC} $ & 
$\sigma_{\rm CC}({\rm GRV98})- \sigma_{\rm CC}({\rm CTEQ6})$ 
  \\  \hline
$5$   &  $5.6 \%$ & $-3.6 \%$      \\
$10$  &  $3.3 \%$ & $-2.5 \%$      \\
$20$  &  $2.8 \%$ & $-1.0 \%$      \\
$50$  &  $2.4 \%$ & $\pm 0.0 \%$   \\
$100$ &  $2.2 \%$ & $+ 0.5 \%$     \\
\end{tabular}
\caption{\label{tab:errors}
Propagation of PDF uncertainties into the evaluation of
$\sigma_{\rm CC}({\nu_{\tau}}N)$ with
a cut $W>1.4\ {\rm GeV}$.  
The second column $\Delta \sigma_{\rm CC}$ was calculated using the CTEQ6 eigenvector
PDFs along the master formula (3) in {\protect \cite{Pumplin:2002vw}}.
The third column compares 
GRV98 {\protect \cite{Gluck:1998xa}} with the central CTEQ6M set.}
\end{table}


\end{document}